\documentclass[11pt]{article}

\usepackage{graphicx}
\usepackage[top=1in,bottom=1in,left=1in,right=1in,paper=a4paper]{geometry}
\usepackage{lineno}
\usepackage{chemfig}
\usepackage{chemformula}
\usepackage{amsmath,bm}
\usepackage{amssymb}
\usepackage{wasysym}
\usepackage{braket}
\usepackage{xcolor}
\usepackage{caption}
\usepackage{subcaption}
\usepackage[normalem]{ulem}

\usepackage[style=nature]{biblatex}
\usepackage{multicol}
\usepackage{float}

\usepackage{hyperref}

\title{Revisiting Claims of Extracranial Biophoton Detection from the Human Brain}
\author{{ \normalsize Vahid Salari\textsuperscript{1,2,}\thanks{These authors contributed equally to this work.} , Vishnu Seshan\textsuperscript{1,2,3,}\footnotemark[1] , Rishabh Rishabh\textsuperscript{1,2,3,}\footnotemark[1]} , \\ {\normalsize Daniel Oblak\textsuperscript{1,2}, Christoph Simon\textsuperscript{1,2,3,}\thanks{Corresponding author: Christoph Simon, Department of Physics and Astronomy, University of Calgary, Calgary, Alberta, Canada. Email: csimo@ucalgary.ca}
}\\\\
{\footnotesize\textsuperscript{1}Department of Physics and Astronomy, University of Calgary, Calgary, Alberta, Canada}\\
{\footnotesize\textsuperscript{2}Institute for Quantum Science and Technology, University of Calgary, Calgary, Alberta, Canada}\\
{\footnotesize\textsuperscript{3}Hotchkiss Brain Institute, University of Calgary, Calgary, Alberta, Canada}\\}

\date{}

\begin{document}
\maketitle


\section*{Abstract}
Ultraweak photon emission, also referred to as biological autoluminescence or biophoton emission, is the spontaneous emission of extremely low levels of light from a broad range of biological systems. Recent studies have reported that UPE measured extracranially can serve as a potential non-invasive biomarker of brain activity. Here, we show that this interpretation suffers from serious problems. We show that, when observed under properly dark conditions, the UPE from the head is much weaker than what is reported in certain papers on `brain UPE' from human heads. We also show that the large signals reported in these studies can be  {explained by} background light contamination. Furthermore, photons with wavelengths $<600$~nm are strongly attenuated by scalp and skull tissues, and longer wavelengths fall largely outside the effective spectral sensitivity of the photomultiplier tubes (PMTs) used. As a consequence, even if UPE from the head is detected under properly background-free conditions, it is likely to be dominated by emission from the scalp rather than from the brain, certainly as long as PMTs are used. Our results emphasize the importance of careful experimental design to make genuine progress on this important question.
\vspace{20mm}

Ultraweak photon emission (UPE) is a spontaneous emission of extremely low levels of light  in the spectral range of $200-1000$~nm from  a broad range of biological systems \cite{UPE1963, UPEreview, salari2025imaging}. {UPE emission without any influence of external stressors or stimuli is termed spontaneous UPE. In contrast, UPE emission initiated by various biotic and abiotic stresses and oxidative factors is termed induced UPE. The intensity of spontaneous UPE emission is in the range of tens to several hundreds of photons/cm$^2$/sec, but can increase up to $10^4$ photons/cm$^2$/sec for induced UPE~\cite{UPEreview}.} It is known that reactive oxygen species (ROS) play an important role in the phenomenon of UPE \cite{ROS-UPE} and UPE can serve as an indicator of cellular and organism oxidative status \cite{UPEreview}. Several experiments have indicated a direct relationship between the intensity of UPE and various aspects of neural activity, including oxidative reactions, EEG activity, cerebral blood flow, cerebral energy metabolism, and the release of glutamate \cite{UPE2, UPE3, UPE4, UPE5, UPE6}. For instance, Kobayashi et al. have reported that in vivo imaging of UPE from rat's brain continuously for 6 hours has a dynamics which correlates with theta electrical activity of the brain \cite{UPE6}.  
\par

Casey et al.~\cite{casey2025} present an intriguing study suggesting that UPE signals recorded over the scalp may reflect functional brain states (following a methodology presented in Ref.\cite{starprotocols}), which is similar to an older study by Dotta et al. \cite{Persinger}. We agree with the general consensus in the field that the brain - and any other living tissue - emits UPE as a consequence of metabolic and oxidative processes, and that correlations between UPE and electrophysiological activity, such as EEG, are in principle plausible and scientifically interesting. However, establishing such correlations requires rigorous optical, spectral, and noise-control considerations. In this work, we show that the experimental methodology and signal interpretation presented in previous studies \cite{casey2025, starprotocols, Persinger} do not adequately meet these requirements, and therefore do not provide a scientifically sound basis for attributing the reported signals to brain-derived UPE.
\par

 \begin{figure}
    \centering
    \includegraphics[width=0.9\textwidth]{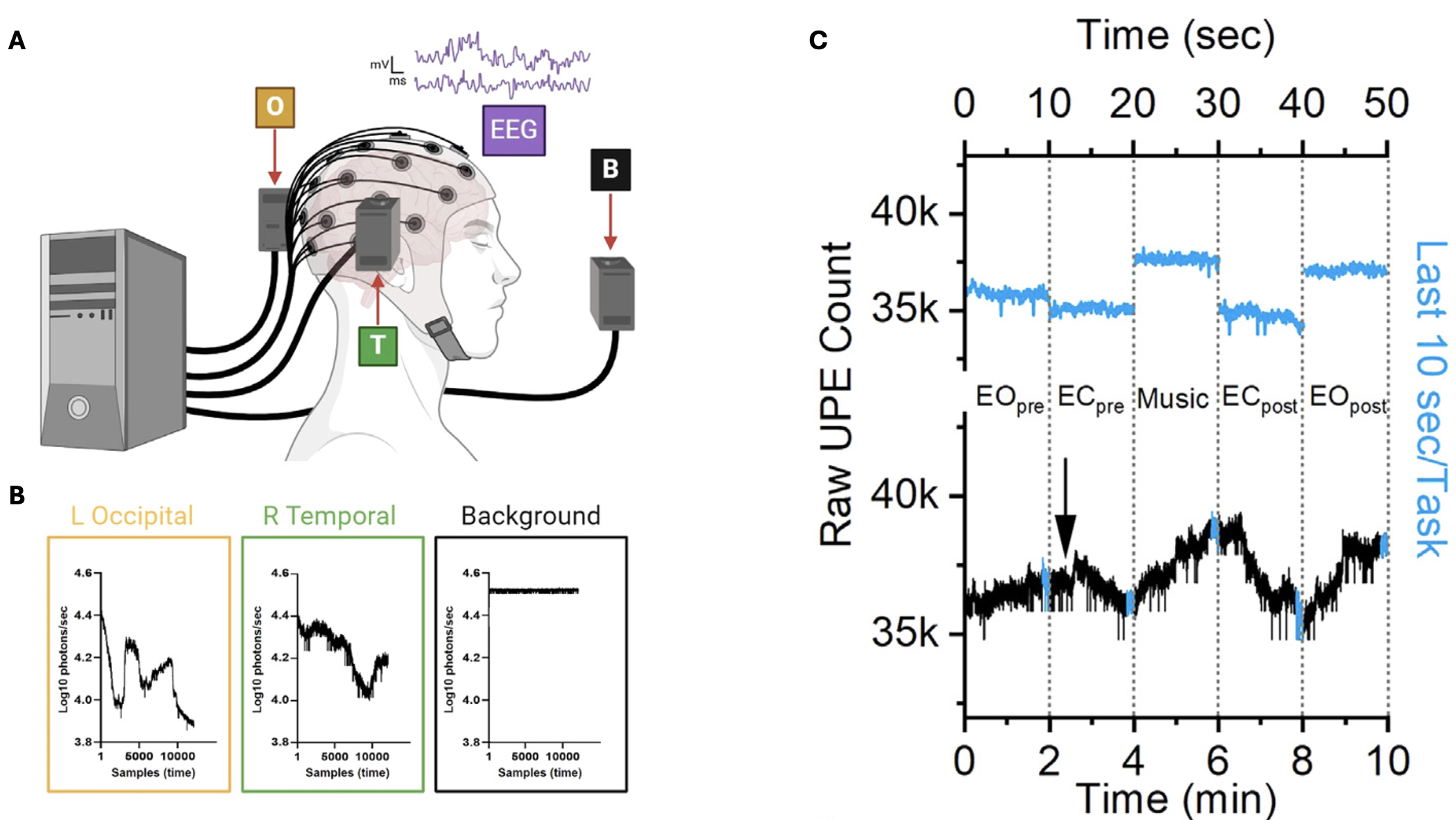}
    \caption{(A) Schematic of Casey et al.'s \cite{casey2025} experimental setup. (B) PMT recordings from left(L) Occipital and right(R) Temporal of the head, as well as the background, reported by Casey et al. (C) PMT recordings from the head, reported by Casey et al. The reported photon counts are much higher than the known UPE intensity from all examined living systems, which are in the range 10-10$^3$ counts per cm$^2$ per sec~\cite{UPEreview}. Moreover, the signal from the head is lower than the background. These figures are reproduced from Ref.~\cite{casey2025}, which is available under a \href{https://creativecommons.org/licenses/by-nc/4.0/}{CC BY 4.0 license}.}
    \label{fig:casey}
\end{figure}

\begin{figure}
    \centering
    \includegraphics[width=0.8\textwidth]{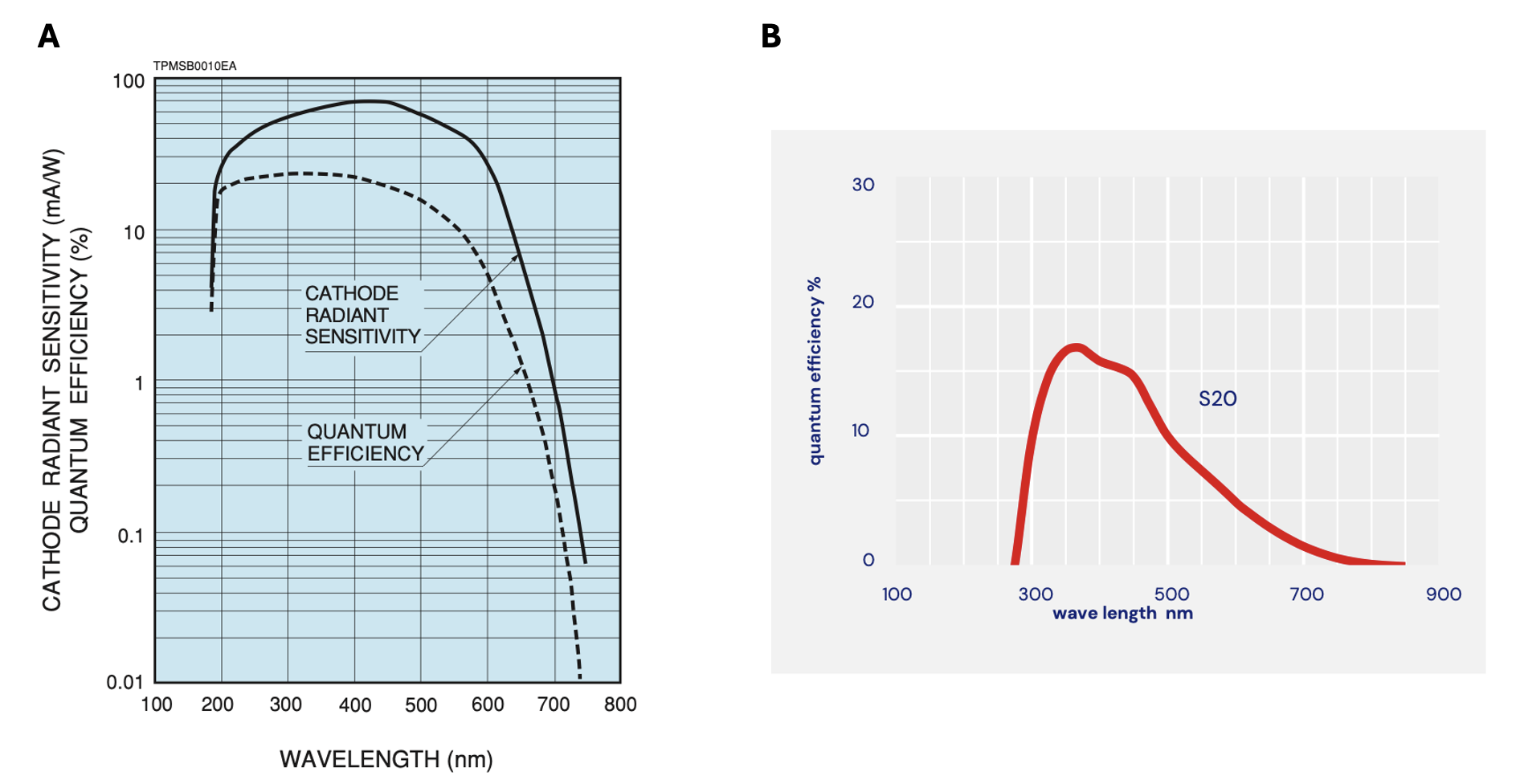}
    \caption{The quantum efficiency diagram of (A) the PMT (Hamamatsu R4220-P) we used in our experiment and (B) the PMTs (SENS-TECH DM0090C) that Casey et al.\cite{casey2025} have used in their study. Note that the quantum efficiencies at wavelengths above 600 nm fall below 5\%. The diagrams are taken from the respective data sheets~\cite{PMT2,PMT1}.}
    \label{fig:pmt}
\end{figure}

Casey et al. reported a count rate of 
$ C \approx 2.5\text{-}4.0 \times 10^{4}\text{ photons/sec} $ 
(see Fig.~\ref{fig:casey}B-C), using a PMT with a photocathode active diameter of $ D = 2.2\,\text{cm} $ and a maximum quantum efficiency of 
$ \mathrm{QE} \approx 0.17 $ (see Fig.~\ref{fig:pmt}B)~\cite{PMT1}. 
Therefore, the resulting photon flux ($F$) at the detector is
\begin{equation}
    F = \frac{C}{\mathrm{QE}} \cdot \frac{4}{\pi D^2} 
\approx 4\text{-}6 \times 10^{4}\,\text{photons/cm}^{2}\text{/sec}.
\end{equation}
\par

The photon flux at the detector can never exceed the flux at the sample surface. However, the photon flux at the detector of Casey et al. is at the higher end of the range reported for induced UPE under strong oxidative stress~\cite{UPEreview}. Together with the fact that their reported background light level is higher than their reported signal (see Fig. 1B), this raises the question whether the detected signal might be dominated by something other than UPE.
\par

As a first step in investigating this issue, we performed an experiment to measure the UPE from a human head under properly dark conditions. The experimental enclosure consists of a light-tight tent within a dark room. The dimensions of the dark enclosure are 2.25m$\times$3.10m$\times$2.15m. A human participant was seated inside the fully enclosed, opaque dark tent placed within a dark room, conditions similar to those of Salari et al.~\cite{salari2025imaging}. A broadband PMT (Hamamatsu R4220-P) was used in this experiment; the quantum efficiency diagram is shown in Fig.~\ref{fig:pmt}A. Note that the response curves of our PMT are similar to those used by Casey et al. (Fig.~\ref{fig:pmt}B). The active photocathode area of our PMT (192 mm$^{-2}$)~\cite{PMT2}, was approximately half the area of the device used by Casey et al (380mm$^{-2}$)~\cite{PMT1}. The PMT was positioned at a fixed distance of approximately 5 cm from the participant's forehead to monitor UPE from the head. This is inspired by Casey et al.'s experiments, in which PMTs were positioned approximately 5 cm above the head surface with their apertures facing the corresponding brain regions of interest. 
\par

\begin{figure}
    \centering
        \includegraphics[width=0.7\textwidth]{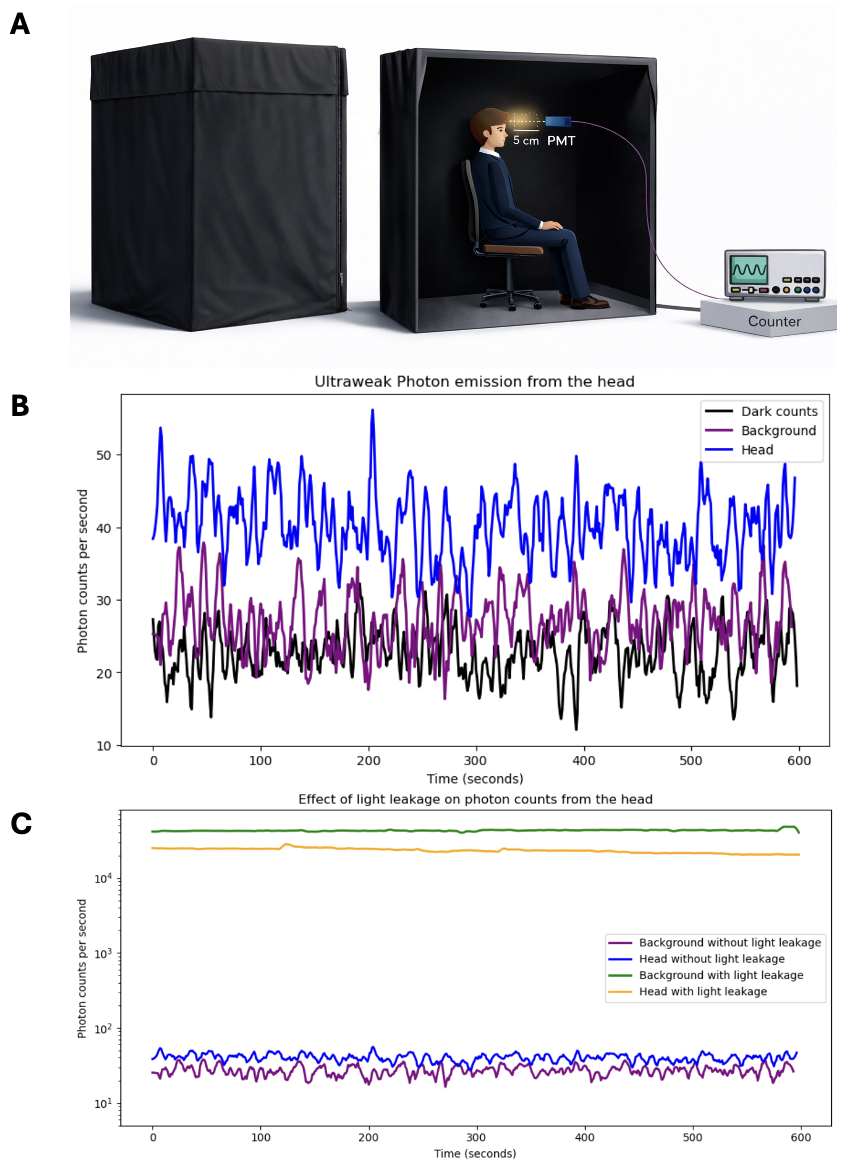}
    \caption{(A) Schematic of our experimental setup, in {which} a participant is seated in a darkened tent while a PMT is positioned at a fixed distance of 5 cm from the forehead to monitor UPE. The lights in the room in which the tent is situated can be turned on or off, and linear apertures of varying sizes can be introduced into the tent cloth. (B) Measurements performed under properly dark conditions. PMT dark count rates{, measured with the detector capped} (black), background {(no human participant)} count rates (purple), and UPE count rates measured from the human forehead (blue) are shown. (C) Measurements performed with optical leakage (10 mm linear aperture with room lights on). Background (no human participant) count rates with optical leakage (green) and photon count rates with optical leakage recorded in the presence of a human participant (orange) are shown. {Similar to Fig.~\ref{fig:casey}, the head signal with leakage is lower than the background signal with leakage. We also show two traces from panel (B) for better comparison: the background (purple) and UPE count rates (blue), both measured under properly dark conditions.}}
    \label{fig:experiment}
\end{figure}

Prior to any measurements on the participant, the intrinsic PMT dark count rate was recorded with the PMT entrance optically capped. This was found to be stable at an average value of approximately 22 counts per second. Then the PMT was uncapped under complete darkness and the background {(no human in the tent)} count rate of approximately 27 counts per second (flux {$\approx 11$} photons/cm$^2$/sec) was observed. {The flux at the detector was calculated using the minimum photocathode effective area of our PMT ($1.92,\text{cm}^2$) and its typical quantum efficiency ($0.23$)~\cite{PMT2}. Furthermore, before calculating the flux, the average dark count rate (22 counts/sec) was subtracted from the measured count rate.} Subsequently, measurements were performed with the PMT uncapped and facing the forehead under complete darkness to record the putative UPE signal. Approximately 40 counts per second were recorded {(measured count rate $-$ dark count rate $=40-22=18$ counts/sec;  flux $\approx 41$ photons/cm$^2$/sec)}. These measurements are shown in Fig.~\ref{fig:experiment}B. \textit{Our putative UPE signal from the head, recorded under complete darkness, is three orders of magnitude lower than the measurements reported by Casey et al. Note that this level is consistent with typical photon fluxes for spontaneous UPE~\cite{UPEreview}.}
\par

This raises the question why Casey et al. observe such large rates. To resolve this we conducted additional experiments to demonstrate that background light contamination can account for the large count rates reported by Casey et al. These measurements are shown in Fig.~\ref{fig:experiment}C. To assess the sensitivity of the measurement to ambient light leakage, controlled leakage tests were performed by introducing a 10 mm linear aperture in the tent fabric with the room lights outside the dark tent switched on. The intensity outside the dark tent under room light-ON condition was roughly ~50 lux, equivalent to approximately $10^{14}$ photons/s/cm$^2$. Under these conditions, the {tent} still appeared completely dark to the naked eye. First, background levels in the absence of a human participant were measured. Background count rates under optical leakage were found to be approximately 40000 counts per second (flux {$\approx 90000$} photons/cm$^2$/sec). \textit{Note that these rates are comparable to the background levels reported in the Casey et al. study.}  We then repeated these experiments in the presence of a human participant, with PMT facing their forehead. Introducing a 10 mm slit with the lights on resulted in the count rates of 25000 counts per second (flux ${\approx 56000}$ photons/cm$^2$/sec). \textit{Note that these measurements match Casey et al.'s reported `UPE' levels.} These observations show that relatively minor light leakage can produce photon detector counts similar to those attributed to brain-derived UPE. As in Casey et al.'s experiment, the head counts that we observe are lower than the background counts, likely because the head blocks some of the background light. Let us note that Casey et al. measured the background signal with the PMT directed away from the participant toward an adjacent wall, whereas our background measurements were acquired using a similar detector geometry and optical configuration as the head measurements, but without a participant present. Despite this slight difference in the configurations, we conclude that the signals which Casey et al. interpreted as `brain UPE' were highly likely to be dominated by background light.
\par

\begin{figure}
    \centering
    \includegraphics[width=0.5\textwidth]{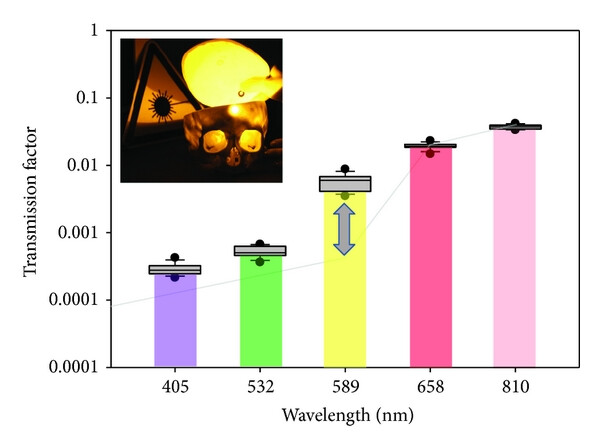}
    \caption{Light transmission in the 405–810 nm wavelength range through human skull bone. This figure is taken from Ref.~\cite{Litscher2014LASERTHERAPIE}, which is available under a \href{https://creativecommons.org/licenses/by/3.0/}{CC BY 3.0 license}.}

    \label{fig:transmission}
\end{figure}


We now show that even the UPE that we recorded under properly dark conditions is highly unlikely to come from the brain. Previous measurements of light transmission through the head show that shorter wavelengths are severely attenuated by skin and skull. Hart et al.~\cite{hart2016new} measured the optical transmission through human skin, skull bone, and brain tissue over wavelengths ranging from 450 to 900~nm (see Fig.~1 of Ref.~\cite{hart2016new}). For scalp skin samples obtained from the crown and the lateral skull, transmission was effectively negligible at wavelengths below 580~nm. A modest increase in transmission was observed between 580 and 640~nm, although transmission values remained below 2\% in this range. Beyond 640~nm, transmission increased sharply, reaching a maximum value around 750~nm. The peak transmission value for the crown skin sample was approximately 24\%, while the lateral skull skin sample showed slightly lower peak transmission. Following the peak, transmission gradually decreased with increasing wavelength, reaching approximately 20\% at 900~nm for both skin samples. For skull samples of varying thicknesses and from different anatomical regions, transmission remained close to zero below 580~nm. An increase in transmission was observed beyond this wavelength. The temporal skull samples exhibited the highest transmission values. The 6~mm temporal skull sample, which showed the highest peak transmission, remained between 11 and 12\% for wavelengths from 710 to 830~nm, reaching a maximum transmission of approximately 12\% near 790~nm before gradually decreasing to approximately 10\% at 900~nm. These transmission results match earlier observations by Litscher et al.~\cite{Litscher2014LASERTHERAPIE}, who measured the transmission factor of laser light at five different wavelengths (405 nm, 532 nm, 589 nm, 658 nm, and 810 nm) through the human skull. They reported that transmission increased with wavelength; however, for wavelengths below 600 nm, it remained less than 1\% (see Fig.~\ref{fig:transmission}). We should note that Hart and Fitzgerald~\cite{hart2016new} used unfixed tissues from two donors: one set consisted of previously frozen samples (skin, skull, brain) that were thawed before measurement, while the other was fresh, unfrozen brain tissue stored at 4 °C overnight and measured the next day. Litscher and Litscher~\cite{Litscher2014LASERTHERAPIE} do not provide detailed information about the tissue condition.
\par

The PMTs used by Casey et al. are Sens-Tech DM0090C, with peak quantum efficiency in the blue--green range and a steep decline towards longer wavelengths (see Fig.~\ref{fig:pmt}A). In practice, typical biophoton PMTs have their highest sensitivity between roughly 350 and 500~nm, show markedly reduced quantum efficiency near 600~nm to 5\%, and are relatively insensitive above $\sim 650$~nm, i.e. below 2\%. Consequently, the detectors are {optimized} for wavelengths that cannot realistically pass through the scalp and skull. They are {insensitive} to the very wavelengths (red / near-infrared) that can, in principle, traverse these tissues with non-negligible transmission. Even if a tiny fraction of $>650$~nm photons escape the head, they fall into the least sensitive region of the detector response. 
\par

\begin{figure}
    \centering
    \includegraphics[width=0.6\linewidth]{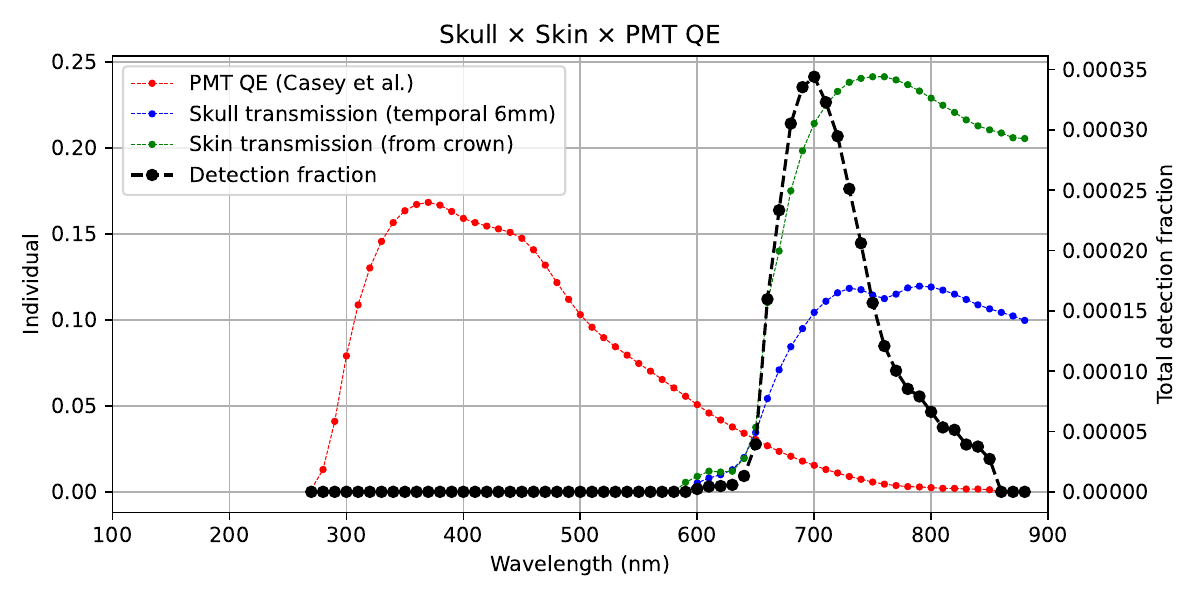}
    \caption{Extracted data for PMT quantum efficiency (Fig.~\ref{fig:pmt}A), skull transmission (Fig.~1B of Ref.~\cite{hart2016new} - temporal, 6 mm) and skin transmission (Fig.~1A of Ref.~\cite{hart2016new} - crown), along with the calculated detection fraction (Quantum efficiency $\times$ Skull transmission $\times$ Skin transmission).}
    \label{fig:detection_fraction}
\end{figure}


We extracted Casey et al.'s PMT quantum efficiency at 10~nm intervals from Fig.~\ref{fig:pmt}A. Similarly, we extracted transmission data for the human skin (crown) and skull (temporal, 6~mm) from Fig.~1A,~B of Ref.~\cite{hart2016new}. The data were extracted using a Python code written with the help of a generative AI tool (ChatGPT Instant 5.3). The detected fraction of brain UPE was estimated as the product of the PMT quantum efficiency and the skull and skin transmission. The wavelength best suited for extracranial detection of `brain UPE' is 700~nm, corresponding to a quantum efficiency of 0.015, skull transmission of 0.104, and skin transmission of 0.214, yielding a peak detection fraction of approximately 0.00034. The calculated detection fraction, along with the extracted data, is shown in Fig.~\ref{fig:detection_fraction}. Casey et al. reported a count rate of 
$2.5\text{-}4.0 \times 10^{4}\text{ photons/sec} $ 
(see Fig.~\ref{fig:casey}B-C), using a PMT with a photocathode active diameter of $2.2\,\text{cm} $, hence the rate of detected photons per unit area is $7-11\times10^3$ photons/cm$^2$/sec. \textit{To achieve an extracranial signal of this magnitude, the brain would need to emit on the order of at least $10^{7}$ biophotons/cm$^2$/second, which is three to six orders of magnitude higher than the expected levels based on previous UPE studies~\cite{UPEreview}.} Therefore, even {when} photons from the head are recorded under properly dark conditions {(see Fig.~\ref{fig:experiment}B)}, they are much more likely to come from the skin than the brain.
\par

The possibility of detecting UPE from the head raises important questions about the potential of UPE as a functional brain readout. {Our results directly challenge Casey et al.'s claim of detecting UPE from the brain. Their reported signal levels are significantly higher than the typical UPE levels from biological tissues, including the scalp. Our head UPE measurements performed under proper dark conditions yield much lower levels. Our experiments also provide a plausible explanation for why their reported levels are so much higher than ours -- background light contamination.} Therefore, a strict optical blackout is essential to minimize ambient light leakage into the detection system, as even trace stray photons can overwhelm true UPE signals. Further, tissue and detector properties impose strict limits. Scalp and skull block nearly all sub-600~nm photons, and the PMTs used are relatively insensitive above 600~nm. Hence, even {when} photons from the head are recorded under properly dark conditions, they are much more likely to come from the skin than the brain. 
\par

Note that even if some statistical structure is present in Casey et al.'s measured signals, this does not imply detection of photons originating from the brain. Instead, such structure {could be due} to statistical artefacts or arise due to a range of alternative mechanisms, including modulation of detector noise, background light fluctuations, variations in detector efficiency, head motion, or coupling to nearby instrumentation (e.g., EEG electrodes). 

\par
In conclusion, for UPE to provide a non-invasive window into brain physiology, it would require the development of measurement protocols that eliminate background light and detectors sensitive to wavelengths transmitted through the skin and skull. With these safeguards in place, UPE could yet become a biomarker for neuroscience, clinical monitoring, and next-generation brain–computer interfaces.


\section*{Acknowledgement}
This work was supported by the National Research Council (NRC) of Canada through its Quantum Sensing Challenge Program and by the Natural Sciences and Engineering Research Council (NSERC) through the Alliance Quantum Consortia Grant ``Quantum Enhanced Sensing and Imaging'' and through its Discovery Grant program.

\section*{Author contributions}
Vahid Salari and Vishnu Seshan performed the experiments. Vahid Salari, Vishnu Seshan, and Rishabh Rishabh carried out the analysis. Vahid Salari, Daniel Oblak, and Christoph Simon supervised the research. Vahid Salari, Rishabh Rishabh, and Vishnu Seshan wrote the manuscript with input from Daniel Oblak and Christoph Simon.

\section*{Competing interest statement }
V.S. is a co-founder and Chief Scientific Officer of vivotraQ Technologies Inc. D.O. serves as a scientific advisor to the company. The remaining authors declare no competing interests. The activities of vivotraQ Technologies Inc. are not related to extracranial monitoring of biophotons from the human head, but instead focus on organ vitality monitoring in transplantation contexts.

\printbibliography

\end{document}